\documentclass{article}
\usepackage{graphicx} 
\usepackage{authblk}
\usepackage{amsmath}
\usepackage{multirow}
\usepackage[sort, comma, numbers]{natbib}
\usepackage{pdflscape}
\usepackage[skip=5pt plus1pt, indent=40pt]{parskip}

\usepackage{geometry}
\usepackage{fancyhdr}
\title{Development of Large Annotated Music Datasets using HMM-based Forced Viterbi Alignment}
\author[1]{S. Johanan Joysingh}
\author[1]{P. Vijayalakshmi}
\author[1]{T. Nagarajan}
\affil[1]{Sri Sivasubramaniya Nadar College of Engineering, Chennai}
\date{}

\usepackage{titling}
\begin{document}
\maketitle

\begin{center}
\section*{Abstract}
\end{center}
Datasets are essential for any machine learning task. Automatic Music Transcription (AMT) is one such task, where considerable amount of data is required depending on the way the solution is achieved. Considering the fact that a music dataset, complete with audio and its time-aligned transcriptions would require the effort of people with musical experience, it could be stated that the task becomes even more challenging. Musical experience is required in playing the musical instrument(s), and in annotating and verifying the transcriptions. We propose a method that would help in streamlining this process, making the task of obtaining a dataset from a particular instrument easy and efficient. We use predefined guitar exercises and hidden Markov model(HMM) based forced viterbi alignment to accomplish this. The guitar exercises are designed to be simple. Since the note sequence are already defined, HMM based forced viterbi alignment provides time-aligned transcriptions of these audio files. The onsets of the transcriptions are manually verified and the labels are accurate up to 10ms, averaging at 5ms. The contributions of the proposed work is two fold, i) a well streamlined and efficient method for generating datasets for any instrument, especially monophonic and, ii) an acoustic plectrum guitar dataset containing wave files and transcriptions in the form of label files. This method will aid as a preliminary step towards building concrete datasets for building AMT systems for different instruments.

\vspace{0.25cm}
\begin{center}
\textbf{Keywords}: \textit{dataset creation, acoustic guitar, automatic music transcription}
\end{center}

\section{Introduction}\label{intro}
Machine learning tasks especially those that use neural networks, rely heavily on the training dataset. In AMT literature, the most common techniques include using signal processing, non-negative matrix factorization and neural networks \cite{b_benetos1}. While signal processing techniques require the least amount of data, with data just enough for experimentation being enough, the Neural Network based systems require huge amount of data to train proper models. In this scenario, there is considerable value for fully and accurately annotated datasets. 

The final aim of AMT would be to transcribe polyphonic(multiple tones played at the same time) multi-instrument music. To achieve this from a wholistic point of view, many efforts have been made to create datasets of polyphonic music. In \cite{b_su} a trained musician plays all parts of a musical piece using different tones on a digital music keyboard. The musician uses his/her experience as a concert musician to play the different parts (instruments) of the music, by ear, or through score sheets if available. MIDI (Musical Instrument Digital Interface) output from the keyboard is extracted and used as new transcription. The problems with this method are that i) the parts (violin, cello, flute, etc.) are synthesized from a digital keyboard, hence losing naturalness of musical instruments themselves ii) playing by listening/ear does not lead to perfectly accurate transcriptions. Multi-track method is used in \cite{b_duan}, where each instrument track is annotated separately, hence making the problem of multi-track annotation into a uni-track variant, which are then combined/mixed in different ways to produce different tracks with annotations. The Bach10 dataset was produced using this method. The drawback of all these methods is that they require good musicians to play or annotate the dataset.

Using a synthesizer/digital keyboard is very common when creating a database, but the drawback is that, we lose essential information about the timbre of real instruments. Hence using data acquired in this fashion for training Neural Networks would lead to inaccurate results for real world signals. The MAPS Dataset as mentioned in \cite{b_emiya} uses the AutoPiano method where audio is generated from MIDI through an electronic keyboard. The TRIOS dataset mentioned in \cite{b_benetos3} also uses a synthesizer to generate audio from MIDI. The advantage of using audio synthesized from MIDI is that, by processing the MIDI files we get perfect annotations/ground truth. These methods, and many others, also use existing pieces of music. This can be disadvantageous in that, we do not have full control of the contents of the dataset. This can lead to inaccurate results as well because, our models rely on what the pieces of music in the training dataset offers, rather than the researcher having complete control over what is being modeled. This results in a bias in the models. Only way to avoid this would be to increase the amount of data used to model, but sometimes this may not be practical. In \cite{xi2018guitarset}, an hexaphonic pickup is used to pick the audio of each string separately, hence simplifying a polyphonic annotation task into individual monophonic ones. This work focusses on the development of a guitar dataset. A drawback of this method is that the annotation is performed semi-automatically with rough annotations being obtained automatically, followed by a manual verification and a correction step. During manual correction, false positives are removed and onset locations are adjusted as required.

Synchronization or alignment of audio with score or audio with audio, using Dynamic Time Warping (DTW) and Dynamic Programming (DP) are common in literature. In \cite{b_ewert}, MIDI-Audio Synchronization is achieved by using DTW on the similarity matrix formed using features developed by the authors and named as Decaying Linearly Normalized Chroma Onset (DLNCO) features. The work is aimed at musical recordings with prominent onsets, as in a piano, while it is meant to not fully collapse in other cases as well. In \cite{b_hu}, MIDI is directly mapped to chromagram vectors instead of synthesizing audio and then generating chromagrams from them. In this case, the timbre of the instrument is considerably ignored. Here too, DTW is performed on a similarity matrix formed using Euclidean distance between the chromagram vectors generated from audio and those assigned to the MIDI, in order to synchronize the two. In \cite{b_turetsky}, audio and audio generated from MIDI files are synchronized using dynamic programming techniques where, the synchronization is done in two levels. First, to obtain the higher order structure and then secondly, to obtain the fine grain details. In \cite{b_joder}, we see yet another technique where both audio and score are converted to chroma features and alignment is performed on them.

In the proposed method, we define note sequences that must be played. These note sequences are similar to exercises that must be practiced when learning a guitar and are easy for an intermediate level musician to play them. This eliminates the need for a very highly skilled musician and also gives complete control over the contents of the dataset, since we define the note sequences or musical phrases. The disadvantage is that, the exercises need to be composed or formulated in the first place. Here we rely on simple patterns that would cover many different note contexts and intervals. The note sequences are played using an acoustic plectrum guitar and are recorded. Since these note sequences extending from 5 to 15 notes are pre-defined, forced viterbi alignment is performed on the audio using them and the note models. These are rough models created with only five examples of a particular note. The details of the process are explained in Section \ref{proposed_method}.

The details of what constitutes an ideal musical dataset, and how this is achieved in the current work are detailed in section \ref{datasets}. The scope of the dataset in the current scenario is also detailed in this section. The general outline of the proposed method and the details of the steps involved including the experimental setup are mentioned in section \ref{proposed_method}. The experiments run and the results obtained are detailed in section \ref{results}.

\section{Datasets}\label{datasets}

Datasets are the centerpiece of machine learning and can make or break a machine learning algorithm. Irrespective of how good the algorithm maybe, if trained with bad data, the end result is a system that produces inaccurate classification or clustering results. Hence focus and attention need to be given in forming the dataset on which machine learning algorithms are trained on. With the availability of highly accurate but data hungry neural network systems, the need for huge datasets with tens of hours of labeled audio, and efficient ways to create them, are all the more justified \cite{b_benetos2}\cite{b_bello}.

A good musical dataset has the following characteristics \cite{b_su}:
\begin{itemize}
    \item Generality: Contains a good representation of real world instrument(s)
    \item Efficiency: Quick and easy to collect, requires less domain knowledge and is not labor intensive
    \item Cost: Does not cost excessive time and money to record and annotate the dataset
    \item Quality: Quality of the recorded audio and the accuracy of the transcription is not compromised
\end{itemize}

Many previous works, as listed in the section \ref{intro}, including \cite{b_emiya} \cite{b_benetos3} have been created using synthesizers, hence losing generality. Although they shine in terms of efficiency, cost and quality of the annotations, they fall short due to the fact that they fail to represent the instrument in its real form. If datasets are generated manually and with real instruments, they have better generalization capabilities, but may lack efficiency and cost more time and money. In the proposed approach we try to balance the amount of work done manually and automatically to create an efficient frame work for dataset generation.

\subsection{The Approach}\label{approach}
One way to look at the approaches taken to create a dataset is to view them as a spectrum of approaches where at one end, every aspect of the dataset is determined and created from scratch(audio and score) and the other end, where existing audio and midi files are processed to generate an annotated dataset. Many methods that exist in literature can be located at some point in this spectrum. For example, some may use existing audio and midi files available freely, and work on synchronization approaches. Some may use existing MIDI files, but create the audio files \cite{b_emiya}, or vice versa. Some may use existing scores of music and play them to create the audio and midi files simultaneously \cite{b_su}. In the proposed approach, we position ourselves near the first end of the spectrum where almost every aspect of the dataset is determined and created from scratch. This means that the notes to be played in each audio file, the audio to be recorded, and the annotations, are all created, performed, recorded and verified by the authors. At first this may seem to be a daunting task, but the reason for going with this approach is manifold, and they are listed below, 
\begin{itemize}
    \item Bias Control: When using audio and midi that exist already, the data maybe biased, in that it might be confined to a particular scale, genre or expression. When the data is composed from scratch, we obtain complete control of the dataset.
    \item Ease of Score Creation: If what is to be played and recorded are confined to simple musical exercises, the process becomes more logical and less creative, enabling spreadsheets and programming languages to compose the note sequences.
    \item Ease of Score Alignment: When the note sequences are composed, the score alignment problem is simplified as will be explained later in section \ref{proposed_method}. Also, the score is simple and does not need to be checked, as is commonly done when using MIDI files to create annotations.
\end{itemize}

\subsection{Scope of the Dataset}\label{scope}
The scope of the dataset created is to train machine learning systems (statistical or neural networks) for automatic transcription of plectrum guitar solo performances. Although the system trained with this dataset can provide really good results when one note is played at a time (monophonic), it can fail in polyphonic cases, for example when the audio contains chords. The primary focus in this work then is not the dataset itself, but rather on what the technique can achieve when used on monophonic instruments like the woodwind and bowed string family, promoting further research in these fields. A guitar is used in this case as a means to test the efficiency of the dataset creation process. The authors believe that, in the task of transcribing multi-track audio, once the tracks (sources) of a multi-track audio file are separated via source separation methods, methods adapted and developed for transcribing individual instruments can be used to obtain accurate transcriptions more efficiently.

\section{Proposed Method}\label{proposed_method}

\subsection{General Outline} \label{general_outline}
The proposed method is an incremental form of dataset generation that works in stages, where the note sequences are made to evolve in complexity as more data is gathered, annotated and the models using the gathered data are formed, hence enabling easier annotation of complex note sequences. As data increases, complex note sequences can be formed in the later stages, and the system will have the ability to align them accurately because the generalization ability of the models also grows with data. For example, in the first stage, just one string of the guitar, say the first string, with just the first five frets in that string are chosen (E4, F4, F4\#, G4, G4\#) and note sequences are generated using these notes. Each of these note sequences are played and recorded multiple times. Models are created with just a few files and then many files that contain the same notes but in different sequences are time-aligned using HMM based forced viterbi alignment technique. This is repeated for all the strings of the guitar, covering just the first four/five frets depending on the string, hence covering the note range from E2 to G4\#. A table representing a guitar fret board, showing the notes that are focused(bolded) in the current work, is shown in Table \ref{tab1}. In all, 29 notes are covered.

\begin{table}[h!]
  \caption{Notes on the guitar fretboard that are covered in the dataset}
  \vspace{0.25cm}
  \begin{center}
    \begin{tabular}{|c|l|l|l|l|l|l|l|}
    \hline
    \textbf{String\textbackslash{}Fret} & \textbf{6} & \textbf{5} & \textbf{4}             & \textbf{3}             & \textbf{2}             & \textbf{1}             & \textbf{0}           \\ \hline
    \textbf{6}                          & A3\#       & A3         & \textit{\textbf{G2\#}} & \textit{\textbf{G2}}   & \textit{\textbf{F2\#}} & \textit{\textbf{F2}}   & \textit{\textbf{E2}} \\ \hline
    \textbf{5}                          & D3\#       & D3         & \textit{\textbf{C\#3}} & \textit{\textbf{C3}}   & \textit{\textbf{B3}}   & \textit{\textbf{A3\#}} & \textit{\textbf{A3}} \\ \hline
    \textbf{4}                          & G3\#       & G3         & \textit{\textbf{F3\#}} & \textit{\textbf{F3}}   & \textit{\textbf{E3}}   & \textit{\textbf{D3\#}} & \textit{\textbf{D3}} \\ \hline
    \textbf{3}                          & C4\#       & C4         & B4                     & \textit{\textbf{A4\#}} & \textit{\textbf{A4}}   & \textit{\textbf{G3\#}} & \textit{\textbf{G3}} \\ \hline
    \textbf{2}                          & F4         & E4         & \textit{\textbf{D4\#}} & \textit{\textbf{D4}}   & \textit{\textbf{C4\#}} & \textit{\textbf{C4}}   & \textit{\textbf{B4}} \\ \hline
    \textbf{1}                          & A5\#       & A5         & \textit{\textbf{G4\#}} & \textit{\textbf{G4}}   & \textit{\textbf{F4\#}} & \textit{\textbf{F4}}   & \textit{\textbf{E4}} \\ \hline
    \end{tabular}
  \label{tab1}
  \end{center}
\end{table}

\begin{figure}[htpb]
  \centerline{\includegraphics[width=0.45\textwidth]{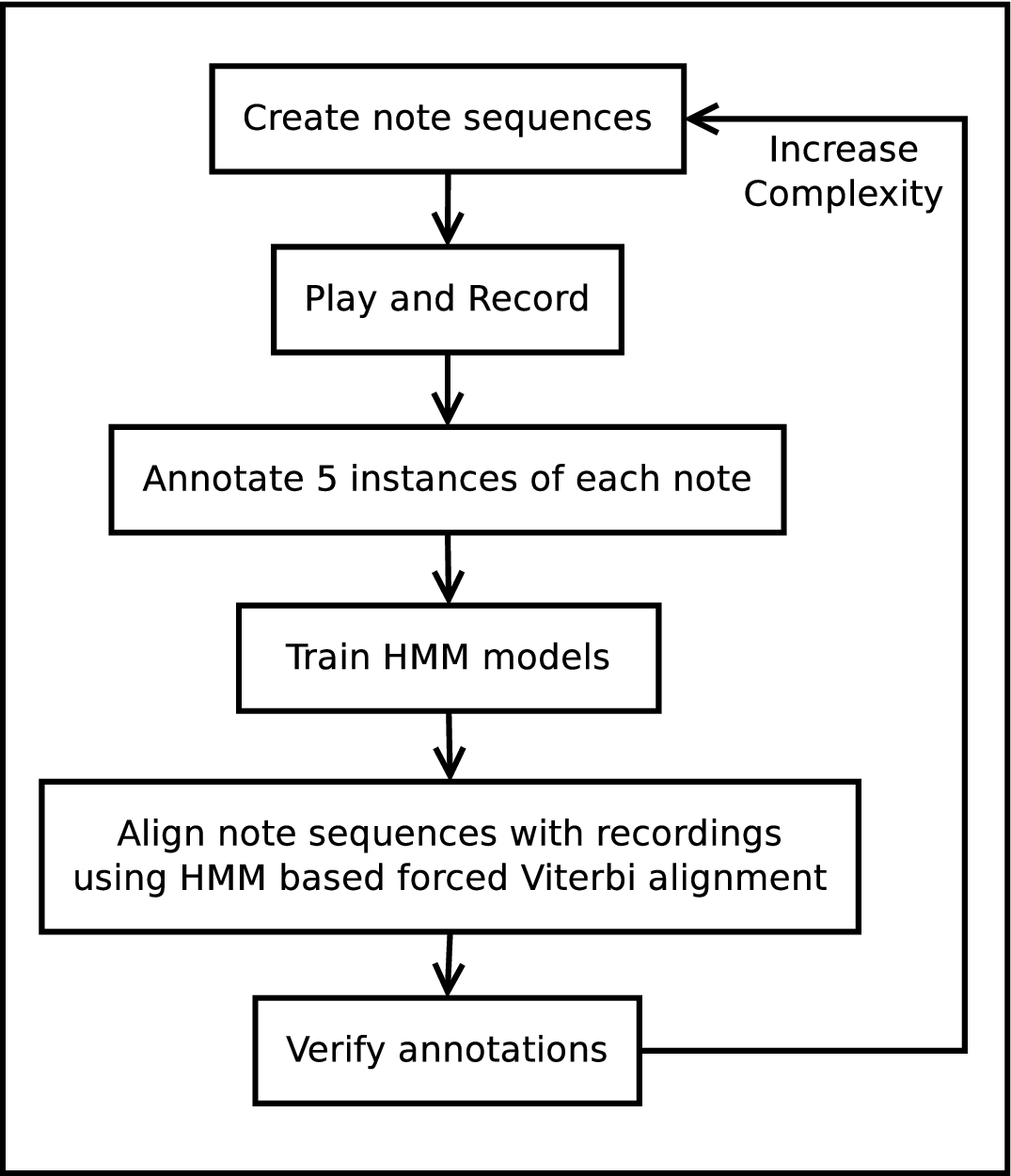}}
  \caption{Generic steps involved in creating a dataset incrementally}
  \label{fig_flow}
\end{figure}

The models generated in this first step can be used in the second stage. Although not part of the current work, in the second stage, note sequences that span over multiple strings, namely all the types of scales (major, minor, pentatonic, etc.) can be used. This process can be extended to few more stages where more note intervals and more of the higher frets can be covered. Hence covering all note intervals and all regions in the guitar. This generic process is outlined in Fig. \ref{fig_flow}.

\subsection{Steps} \label{steps}
The proposed method that is used to generate models for the notes ranging from E2 to G4\# can be summarized in the following steps,
\begin{itemize}
  \item Select string
  \item Determine note sequences for the string, such that it covers the first four/five frets
	\begin{itemize}
		\item example note sequence on the first string: E4, F4, G4, E4, F4\#, G4\#, E4, G4\#, F4\#, E4, G4, F4
	\end{itemize}
  \item Play each note sequence multiple times on the acoustic guitar and record the audio
  \item Annotate a few files, such that a minimum of 5 instances of each note is covered
  \item Build a HMM Model with a single state and single mixture component, using 39 dimensional MFCC features which include 13 each of velocity and acceleration coefficients 
  \item Use the HMM Models to time-align the audio files using forced viterbi alignment technique
  \item Repeat the same process for the next string until all six strings are covered
\end{itemize}

\subsection{Experimental Setup}
The recordings are performed in a studio environment with a Rode NT1-A condenser microphone and a 41 inch acoustic guitar made by Vault. The audio files are recorded at a sampling rate of 44,100 Hz and bit depth of 16 bits and saved in WAV format. The labeling of the audio files is done using wavesurfer and the labels can be accessed via a text editor as well.

\begin{figure}[h]
  \centerline{\includegraphics[width=0.80\textwidth]{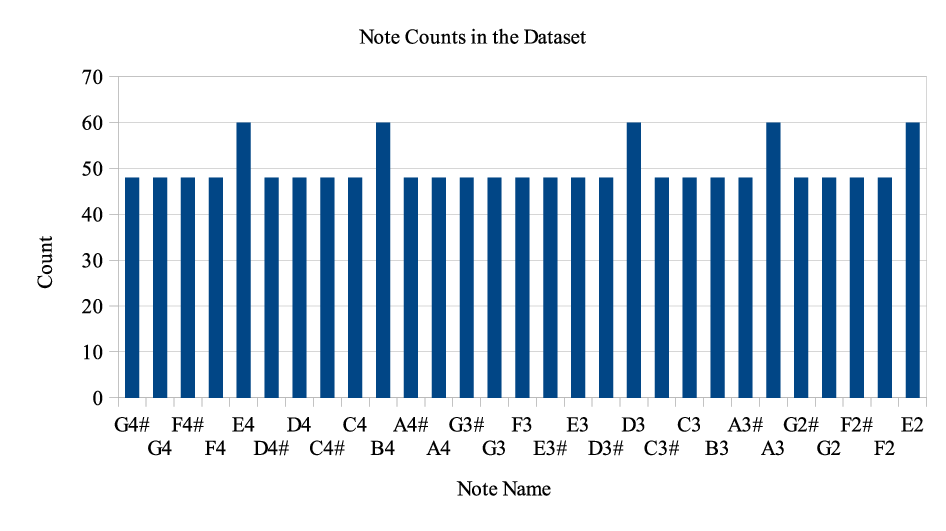}}
  \caption{Count of each note in the dataset}
  \label{fig_chart}
\end{figure}

In the dataset created, three note sequences are formulated for each of the six strings of the guitar. Each of these note sequences are recorded 12 times. In total, there are 216 wave files present in the dataset. This is split into 36 wave files for each string. Each of the 29 notes in the dataset is approximately played 50 times. A bar chart showing the counts of each of these notes, is shown in Fig. \ref{fig_chart}.

For HMM based forced alignment, 5 wave files are considered for each string. These files contain the note sequence that has all the notes selected in the string, hence enabling us to create models for all the notes selected in the string using these five wave files. These wave files are manually labeled and annotated. HMM models with a single state and single mixture component are created from these files. These models and the note sequences are then used to annotate the rest of the 31 files recorded for each string. But there is absolutely no limit as to how many files can be time-aligned by this method once the initial five files are labeled manually. This is provided that, i) the note sequences are available and ii) the note sequences only contain notes for which the models are already built in the previous step.

\section{Results and Contributions} \label{results}
\subsection{Results}

\begin{figure}[h!]
  \centerline{\includegraphics[width=0.95\textwidth]{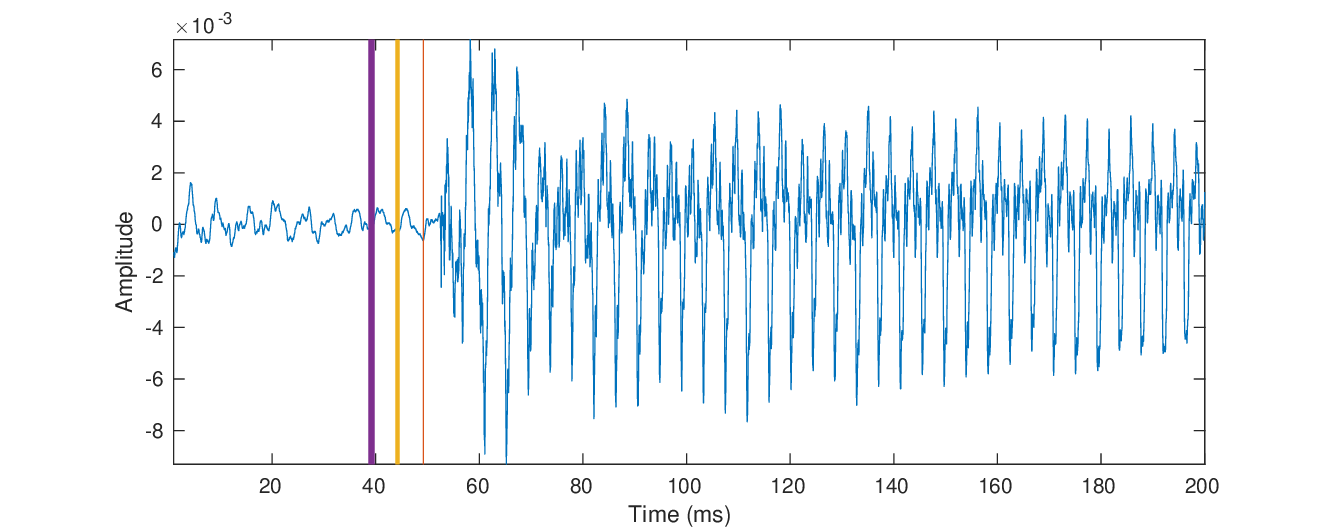}}
  \caption{Time domain representation of the onset region of a note, showing the actual onset location and onset locations with error of 5 and 10 ms}
  \label{fig_onsets}
\end{figure}

Since the dataset is created using an acoustic instrument, the annotations are verified manually. The verification is carried out by a person with experience in music accompaniment, composition and production. The maximum error from the onset location is 18ms and the average error is 7ms. These errors are usually constant for a particular file and are easily rectified by shifting the labels programatically by a few milliseconds. Doing this yields better results, with a maximum onset location error of less than ~10ms and an average of ~5ms. In onset detection literature, the usual threshold of this error even for hand-labelling is 25 to 50ms\cite{b_bello2}. The maximum error obtained in the current work is far less than the acceptable standards. Figure \ref{fig_onsets} show the onset region of a note. The actual onset location is marked by the thinnest vertical line, the thicker lines towards the left represent onset locations that are 5 and 10 ms away from the actual onset location.

The amplitude and frequency characteristics of a single note evolves with time. When annotating notes, if at all there is an error, it is better to have the label marked from a point before the onset of that note, than after. Doing this would help model the whole note accurately for AMT tasks. Hence during correction, more weight is given to errors that caused the labels to exclude the onset.

In order to get an idea of the distribution of error in the dataset, the onset locations are observed and the error in milliseconds is noted. The cumulative distribution of error under each error threshold is shown in Table \ref{tab_results}.

\begin{table}[h!]
  \caption{Cumulative percentage of notes that fall under a certain error threshold}
  \vspace{0.25cm}  
  \begin{center}
    \begin{tabular}{|c|c|}
      \hline
      \textbf{Error (ms)} & \multicolumn{1}{l|}{\textbf{Percentage of Notes}} \\ \hline
      \textbf{2}          & 36.36\%                                           \\ \hline
      \textbf{4}          & 57.58\%                                           \\ \hline
      \textbf{6}          & 68.18\%                                           \\ \hline
      \textbf{8}          & 90.91\%                                           \\ \hline
      \textbf{10}         & 100.00\%                                          \\ \hline
      \end{tabular}
  \label{tab_results}
  \end{center}
\end{table}

\subsection{Contributions}
The contribution is two fold: an efficient methodology to obtain a dataset for any instrument and a dataset containing audio and label files. The methodology can be used on any instrument and works particularly well with monophonic instruments like the instruments from woodwind family and the bowed string family. Since there are open areas of research for these instruments particularly, for example, onset detection, since they have a slow attack, creating datasets using these instruments will promote research in those fields. An average of 50 instances of each note mentioned in Table \ref{tab1} is collected. The verified annotations are available as label files that can be opened on wavesurfer. These contain the beginning and end times of a note.

\section{Conclusion and Future Work}
The current work shows the possibility of easily creating datasets from scratch, rather than working on real world datasets that contain music from albums and orchestral sessions, donated for use in research. The advantage of building a dataset from scratch is discussed. Using real world instruments over synthesized ones is justified by the fact that the synthesized audio does not contain essential information about the timbre of the instrument. The efficiency of the framework is discussed in terms of time and effort required. A new dataset is created and the time-aligned transcriptions are generated using HMM based forced viterbi alignment. The maximum error is less than 10ms. Although the current work requires manual labeling, the effort required is very minimal (only five labels for each note). Once models are built using these labels, exponentially large amount of data can be annotated, which can be used to further build more robust models.

The application of this technique to build datasets for other instruments, especially the monophonic ones, is encouraged. Such datasets would aid in development of research in those areas. As mentioned in section \ref{general_outline}, the current corpus will be expanded by repeating the process mentioned in section \ref{steps} for a few more stages, incrementally.


\end{document}